\documentclass[gmd, manuscript]{copernicus}

\usepackage{bbm}
\usepackage{xcolor}
\usepackage{todonotes}

\nolinenumbers

\begin{document}

\title{Towards accurate extreme event likelihoods from diffusion model climate emulators}


\Author[1][pmanshausen@nvidia.com]{Peter}{Manshausen} 
\Author[1]{Noah}{Brenowitz}
\Author[1]{Julius}{Berner}
\Author[1]{Karthik}{Kashinath}
\Author[1]{Mike}{Pritchard}

\affil[1]{NVIDIA, Santa Clara, CA, USA}

\runningtitle{TEXT}
\runningauthor{TEXT}

\received{}
\pubdiscuss{} 
\revised{}
\accepted{}
\published{}


\firstpage{1}

\maketitle

\begin{abstract}
ML climate model emulators are useful for scenario planning and adaptation, allowing for cost-efficient experimentation. Recently, the diffusion model Climate in a Bottle (cBottle) has been proposed for generation of atmospheric states compatible with boundary conditions of solar position and sea surface temperatures. Crucially, cBottle can be guided to generate extreme events such as Tropical Cyclones (TCs) over locations of interest. Diffusion models such as cBottle work by approximating the probability density of the training data. Here, we show use cases of the probability density estimates of atmospheric states obtained from this climate emulator. Most importantly, these estimates allow us to calculate likelihoods of extreme events under guidance. When guiding the model towards states including TCs, comparing the probability density under the guided and unguided model enables us to quantify how much more likely the guidance has made the TC. We show how these odds ratios allow us to importance-sample from the TC distribution, reducing the standard error of the probability estimate compared to simple Monte Carlo sampling. Furthermore, we discuss results and limitations of the application of model probability densities to extreme event attribution-like experiments. We present these early but encouraging results hoping they will spur more research into probabilistic information that can be gained from diffusion models of the atmosphere.
\end{abstract}

\copyrightstatement{CC-BY} 

\introduction  
Generative Artificial Intelligence has achieved remarkable success across the modalities of natural language, images, and various scientific applications in recent years. One such success is in medium-range weather forecasting \citep{keisler}, matching and then surpassing \citep{fourcastnet, gencast} the forecast skill of numerical weather prediction and being increasingly deployed operationally \citep{aifs}. For weather applications, research is pushing the boundaries of precision \citep{atlas}, increasingly high resolution \citep{stormcast, corrdiff}, data assimilation \citep{healda}, and direct observation prediction \citep{graph-dop, stormscope}. 

For climate, research has focused on stable rollouts of dynamical (auto-regressive) models, which predict a following timestep given a current one. Stability is needed on a much longer timescale than for weather, in order to allow for studies into climate variability. Some models have shown promising skill at emulating the present-day climate \citep{ace2, camulator, dlesym}. Another approach to simulating climate is the direct generation of samples, without a previous time step as input. This way, \cite{climatebench} have been able to emulate the behavior of a climate model, conditioned on anthropogenic emissions. \cite{bouabid2025score} use diffusion models to generate impact-relevant variables sampled from the distributions of three CMIP6 models conditioned on global mean surface temperature. \cite{perkins2026hiroacefastskillfulai} demonstrate downscaling of 100km climate runs to 3km resolution precipitation fields.

Climate-in-a-Bottle (cBottle, \citep{cbottle}) is a recent family of diffusion models that emulate the present-day climate at unprecedented 5km resolution. Conditioned on sea surface temperatures and solar position, the first step coarse model generates full atmospheric states of winds, temperature, humidity, and geopotential at eight pressure levels. Being a diffusion model, it has the unique capability for climate emulators to be guided to states of interest, in a way similar to the implementation of score based data assimilation \citep{rozet} with a diffusion weather emulator by \cite{sda}. \cite{cbottle} demonstrate this by training a classifier of Tropical Cyclone (TC) probability and using it for classifier guidance, allowing a user to generate samples that have a TC in a user-defined location. Such oversampling can be useful to study the impacts of these extremes, for scenario planning, and building intuition for possible events. However, the model does not inform the user about the likelihood of such generated samples. 

For diffusion models, which learn to estimate the probability density function of the training data, there are theoretical approaches to calculating the probabilities of samples. \cite{song2020score} detail how these probability estimates can be obtained from the probability flow (PF) ODE associated with the diffusion model. However, these PDF values can be counterintuitive, with \cite{nalisnick2019detecting} proving that models can assign higher values to samples from data sets other than the training set. 

Here we show that, nevertheless, probabilistic information can be accurately recovered for guided samples of TCs using the cBottle diffusion model. Specifically, we can quantify how much more likely guidance makes us to obtain a TC sample, which can be used to back out the likelihood of the TC under the unguided model. We further demonstrate other applications of the sample probability calculated from cBottle. The remainder of the paper is organized as follows: Section~\ref{sec:methods} details the methods used here, with subsections~\ref{sec:meth-diff} discussing more general concepts and the cBottle emulator, subsection~\ref{sec:meth-is} describing the methods related to obtaining TC likelihoods, and subsection~\ref{sec:methods-cc} detailing the application to ERA5 samples. Section~\ref{sec:results} shows our results for guidance, TC likelihoods, and other applications, before the final section concludes.

\section{Methods}\label{sec:methods}

\subsection{Diffusion models}\label{sec:meth-diff}
Denoising diffusion models \citep{ho2020denoising, song2020score} such as cBottle allow us to sample from the data distribution by iteratively reversing a stochastic noising process. The noising process adds Gaussian noise $\boldsymbol{\epsilon}\sim \mathcal{N}(0, \sigma^{2} \mathbf{I})$ to a sample $\mathbf{x}_0$ of the data, and we train a score-matching network $\mathcal{D}_{\theta}(\cdot; \sigma)$ (the denoiser) to recover the sample with the objective  
\begin{equation}
    \arg\min_{\theta} \mathbb{E}_{\mathbf{x}_0 \sim p_{\text{data}}} \mathbb{E}_{\sigma \sim p_{\sigma}} \mathbb{E}_{\boldsymbol{\epsilon} \sim \mathcal{N}(0, \sigma^{2} \mathbf{I})} \left\lVert \mathcal{D}_{\theta}(\mathbf{x}_0 + \boldsymbol{\epsilon}; \sigma) - \mathbf{x}_0 \right\rVert^{2} .
    \label{eq:diffusionobj}
\end{equation}
The denoiser approximates the score function, the gradient log-probability density function of the noised data:
\begin{equation}
    \nabla_{\mathbf{x}} \log p(\mathbf{x}; \sigma) \approx \frac{\mathcal{D}_{\theta}(\mathbf{x}; \sigma) - \mathbf{x}}{\sigma^{2}}
    \label{eq:scoreapprox}
\end{equation}
We can sample from the data distribution by solving the corresponding ODE
\begin{equation}
\frac{d\mathbf{x}}{d\sigma} =\frac{\mathbf{x}- \mathcal{D}_{\theta}(\mathbf{x}; \sigma)}{\sigma} \label{eq:ODE}
\end{equation}
from a sample of Gaussian noise (latent) $\mathbf{x}_T$ at a high noise level $\sigma_{max}$ to a low noise level $\sigma_{min}$. For this, we use the EDM sampler proposed by \cite{edm}. 

\subsubsection{Probability Flow ODE}
The ODE formulation allows us to back out the model's estimated probability density for individual samples by \citep{song2020score} (Appendix D):
\begin{equation}
    \label{eq:pfode}
    \log p(\mathbf{x}_0) = \log p(\mathbf{x}_T) + \int_{0}^{T} \nabla \cdot {\mathbf{v}}_{\theta}(\mathbf{x}(t), t) \, dt,
\end{equation}
where ${\mathbf{v}}_{\theta}(\mathbf{x}(t), t)$ is short for the right hand side of the ODE in eq.~\ref{eq:ODE}, the flow velocity. This can be understood intuitively: The probability density of the sample $\mathbf{x}_0$ is equal to that of the latent $\mathbf{x}_T$, plus the diffusion time-integrated change in probability density, which by the continuity equation is 
\begin{equation}
\frac{d}{dt}\log p(\mathbf{x}(t)) = -\nabla \cdot {\mathbf{v}}_{\theta}(\mathbf{x}(t), t)
\end{equation}
The divergence can be evaluated using the Skilling-Hutchinson trace estimator
\begin{equation}
    \nabla \cdot {\mathbf{v}}_{\theta}(\mathbf{x}, t)
    = \mathbb{E}_{p(\boldsymbol{\epsilon})}\bigl[\boldsymbol{\epsilon}^{\top} \nabla {\mathbf{v}}_{\theta}(\mathbf{x}, t) \boldsymbol{\epsilon}\bigr],
    \label{eq:trace_estimator}
\end{equation}
with the Jacobian $\nabla {\mathbf{v}}_{\theta}$ and the isotropic random variable $\boldsymbol{\epsilon}$ (see below, section~\ref{sec:implementation} for the implementation).  

Note that eq.~\ref{eq:pfode} yields the log probability density under the model's learned distribution, rather than the true data distribution. We now introduce some elementary definitions to clarify future interpretation and show that the negative log-likelihood is bounded below by the entropy of the data. First off, the log likelihood of the model describes the expectation of the log probability over the data given the model parameters $\theta$, and is given by
\begin{equation}
 \ell(\theta) = \mathbb{E}_{\mathbf{x} \sim p_{data}}[\log p_\theta(\mathbf{x})].
\end{equation}
Diffusion models are trained with a loss function that is a reweighted approximation of $\ell(\theta)$ \citep{ho2020denoising}.

The log-likelihood of the model is composed of the entropy of the data $H(p_{data})=-\mathbb{E}_{\mathbf{x} \sim p_{data}}[\log p_{data}(\mathbf{x})]$ and the Kullback-Leibler divergence between the true and modeled distributions:
\begin{equation}
   -\ell(\theta)= H(p_{data}) + D_{KL}(p_{data} \parallel p_\theta),
\end{equation} 
where the Kullback-Leibler divergence is 
\begin{equation}\label{eq:kl-div}
    D_{KL}(p \parallel q) = \mathbb{E}_{\mathbf{x} \sim p} \left[ \log \frac{p(\mathbf{x})}{q(\mathbf{x})} \right].
\end{equation}
With mild regularity assumptions for $p$ and $q$, $D_{KL}(p \parallel q)\geq0$ with equality only when $p$ and $q$ are identical almost everywhere.

The decomposition above bounds only an expectation over the data distribution, and says nothing about the accuracy of $\log p_\theta(\mathbf{x})$ for an individual sample $\mathbf{x}$, which is what we compute below. To bound the sample-wise error we instead use
\begin{align}\label{eq:sample-bound}
    \mathbb{E}_{p_{data}}\left|\log p_{data}(\mathbf{x}) - \log p_\theta(\mathbf{x})\right| = D_{KL}(p_{data}\parallel p_\theta) + 2\,\mathbb{E}_{p_{data}}\left[\left(\log \frac{p_\theta(\mathbf{x})}{p_{data}(\mathbf{x})}\right)_+\right] 
    \\ \label{eq:fwd_kl}
    \le D_{KL}(p_{data}\parallel p_\theta) + \sqrt{2 D_{KL}(p_{data}\parallel p_\theta)},
\end{align}
where $(\cdot)_+=\max(\cdot,0)$; the second term is bounded by twice the total variation distance using $\log t \le t-1$, and the total variation distance is in turn bounded by $\sqrt{D_{KL}/2}$ via Pinsker's inequality (for an expansion of these steps, see Appendix~\ref{sec:app-pinsker}). Motivated by the connection between denoising score matching and maximum-likelihood training~\citep{song2021maximum}, we can expect a small KL divergence in~\eqref{eq:fwd_kl} (and thus a small expected sample-wise error) after successful training of our model $p_\theta$.

This formalizes that the accuracy of individual sample probabilities calculated here, in an expected-absolute-error sense, depends only on the model having finite KL divergence from the data distribution, and specifically on how well the underlying model approximates it.

\subsection{cBottle}\label{sec:cbottle}
cBottle is a family of diffusion models, consisting in its first version of: a coarse generator at HPX64 (ca. 100km) resolution, a video model at the same resolution, generating 12 time steps at 6h intervals, and a patch-based super-resolution model, increasing the resolution by a factor of 16 to HPX1024. In this work, we use only the coarse generator.

This model is trained as a denoiser of ICON and ERA5 data and conditioned on a map of SSTs and the time of day and day of year, generating atmospheric states compatible with these oceanic and solar boundary conditions. TC guidance logits come from a \emph{separate} checkpoint (\texttt{cBottle-3d-tc}) which outputs a coarse map (HPX8) of TC probabilities for an input of a given noisy sample. This is what we refer to as the classifier below, used for guidance to TC states. Although it is a distinct checkpoint from the state denoiser, this classifier is itself noise-conditioned: it takes the noisy state and noise level $(\hat{\mathbf{x}}, \hat{\sigma})$ as input and is trained jointly as a denoiser with a binary cross-entropy classification head over the same $\sigma$ distribution used for the main denoising objective (eq.~\ref{eq:diffusionobj}). It has therefore seen noisy inputs at every noise level during training. For a more complete discussion of the model, see \cite{cbottle}. 

Over the course of this research, we found that TC statistics of the cBottle model can be improved by using exponential moving average (EMA) weights, rather than cBottle's published raw checkpoints. This is due to checkpoint-to-checkpoint variability in the generation of TCs, which is mitigated by the moving average. It comes at no disadvantage to the other climatologies reported in \citep{cbottle}. The results in the following are all obtained from the EMA version of the checkpoints. 

\subsection{Extreme event guidance}\label{sec:guidance-method}
In cBottle, guidance is implemented by adding a term to the ODE (classifier guidance): 
\begin{equation}
\frac{d\mathbf{x}}{d\sigma} =\frac{\mathbf{x}- \mathcal{D}_{\theta}(\mathbf{x}, \sigma)}{\sigma} + \gamma \nabla_{\mathbf{x}} \mathcal{L}_{BCE}(\mathcal{C}(\mathbf{x}), y).\label{eq:cguidance}
\end{equation}
Here, $\gamma$ represents a guidance scale hyperparameter, and $\mathcal{L}_{BCE}(\mathcal{C}(\mathbf{x}), y)$ the binary cross-entropy loss of the classifier output $\mathcal{C}(\mathbf{x})$ and the guidance input $y$. 

This can be motivated theoretically: We replace the score of $p(\mathbf{x}; \sigma)$ in eq.~\ref{eq:scoreapprox} with $p(\mathbf{x}; \sigma|y)=p(y|\mathbf{x};\sigma) p(\mathbf{x}; \sigma)$ where $y$ is a condition such as `TC in Miami'. The learning objective of the classifier is $-\log p(y|\mathbf{x};\sigma)$, so
\begin{equation}
    \nabla_{\mathbf{x}} \log p(\mathbf{x}; \sigma|y) \approx \frac{\mathcal{D}_{\theta}(\mathbf{x}; \sigma) - \mathbf{x}}{\sigma^{2}} -\nabla_{\mathbf{x}} \mathcal{L}_{BCE}(\mathcal{C}(\mathbf{x}), y)
    \label{eq:scoreapprox_guided}
\end{equation}
Hence, 
\begin{equation}
\frac{d\mathbf{x}}{d\sigma} =\frac{\mathbf{x}- \mathcal{D}_{\theta}(\mathbf{x}, \sigma)}{\sigma} + \sigma \nabla_{\mathbf{x}} \mathcal{L}_{BCE}(\mathcal{C}(\mathbf{x}), y).\label{eq:cguidance_derived}
\end{equation}
This means that we would be sampling from the exact posterior of the distribution of states given a TC if we were to set $\gamma = \sigma$. However, \cite{cbottle} instead use an adaptive, norm-matched guidance scale; we review that scheme and contrast it with ours in Appendix~\ref{sec:adaptive-guidance}.

In this work, we find that setting a constant $\hat{\gamma} \sim 20 \times 64 = 1280 $ consistently produces guided states without the complicated rescaling used by \cite{cbottle}. The factor of 64 is explained by a resolution difference between the state space of the classifier and that of the denoiser: the classifier operates in HPX8 resolution, while the denoiser operates in HPX64. This difference of a factor of 8 means that for the same state the classifier sees 64 times fewer pixels. Therefore, we need to adjust $\gamma=64 \sigma$. As we will show in section~\ref{sec:implementation}, it can be effective and advantageous to guide just for a small window of the denoising process, which is the window where the phenomenon of interest is generated. For TCs, this happens to be around $\sigma=20$. Our final parameterization then is $\gamma = 64 \cdot 20 \cdot \mathbb{I}_I $ where $I=[15,20]$.

\subsection{Importance sampling}\label{sec:meth-is}
We would like to answer questions like: ``How likely is a TC in Miami, Florida, in October?" From cBottle we can already obtain an answer to this question, by just drawing $K$ samples $\mathbf{x}_i$ from the distribution of October atmospheric states, and counting the fraction of times they contain a TC in the pixel of Miami. 
\begin{equation}\label{eq:mc}
    p(TC) \approx \frac{1}{K}\sum_{i=1}^K \mathbb{I}_{TC}(\mathbf{x}_i'), \hspace{20pt} where \ \ \mathbf{x}_i'\sim p_{\mathrm{cBottle}}
\end{equation}
Here, $\mathbb{I}_{TC}(\cdot)$ indicates the presence of the TC in the location of interest. However,  extremes are rare and thus require many inferences $K$ for an accurate estimate. 

We would like to sample extreme events with guidance, but using those states in the above estimator directly would be wrong. We also need to know \emph{how much we have oversampled}, i.e. how much more likely these states are with guidance than without. Then, we can sample with guidance (pushing the model to extremes), but correct by multiplying with the odds ratio of the guided vs. unguided probability.
\begin{equation}
    o(\mathbf{x}) = \frac{p_{\text{unguided}}(\mathbf{x})}{p_{\text{guided}}(\mathbf{x})}
\end{equation}
We fix this sign convention throughout: because guidance makes a TC sample more likely, $p_{\text{guided}}(\mathbf{x}) > p_{\text{unguided}}(\mathbf{x})$ for such a sample and therefore $\log o(\mathbf{x}) < 0$. In terms of the decomposition of eq.~\ref{eq:deltas} below, this means $\Delta_{latent}<0$ and $\Delta_{denoiser}<0$ for a guided TC sample, while the guidance-divergence term as it is plotted in Figs.~\ref{fig:guide-interv} and \ref{fig:lower-guidance-scale}e) is $-\Delta_{guidance}$, which is expected to be positive.
This is the Importance Sampling estimate \citep{biondini2015introduction}:
\begin{equation}\label{eq:is}
    p_{IS}(TC) \approx \frac{1}{K} \sum_{i=1}^K \mathbb{I}_{TC}(\mathbf{x}_i) o(\mathbf{x}_i) , \hspace{20pt} where \ \ \mathbf{x}_i\sim p_{\mathrm{guided}}
\end{equation}
where now the samples $\mathbf{x}_i$ are drawn from the guided distribution.
The variance of this estimate, i.e. of the mean of the $K$ weighted indicator terms, is
\begin{equation}\label{eq:is-var}
        \widehat{\operatorname{Var}}(p_{IS}) = \frac{1}{K(K-1)}\sum_{i=1}^K\left[\mathbb{I}_{TC}(\mathbf{x}_i) o(\mathbf{x}_i) - p_{IS}\right]^2,
\end{equation}
which we report as a standard error of the mean, $\sqrt{\widehat{\operatorname{Var}}(p_{IS})}$. ERA5 frequency error bars are instead obtained by block bootstrapping (4-day blocks, 1000 replicates).

\subsubsection{Guided model in PF ODE}
\begin{figure}
    \centering
    \includegraphics[width=0.75\linewidth]{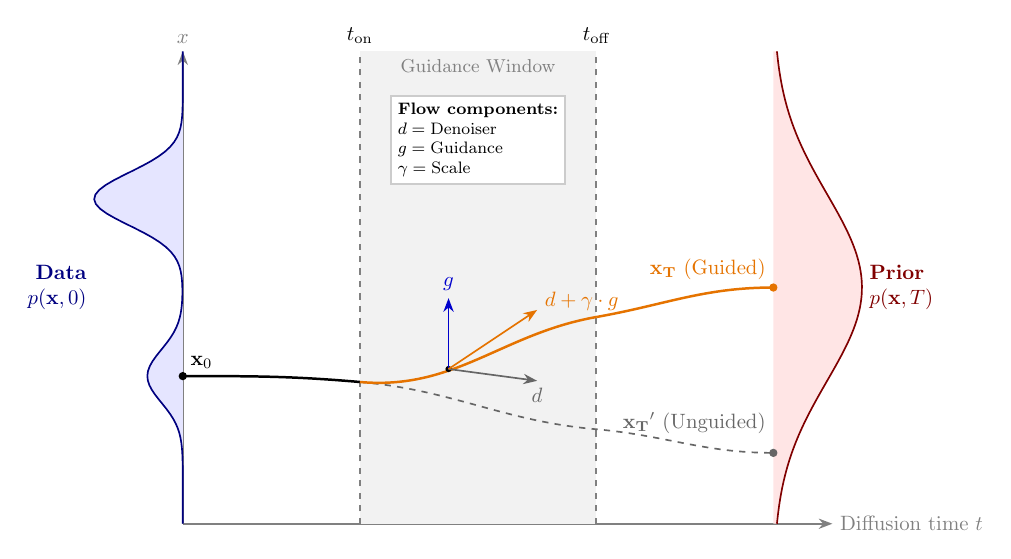}
    \caption{Visualization of the PF ODE. To sample from the data distribution, we would draw a latent from the prior on the right and flow along the ODE velocity to the left. For calculating the probability of a sample, we do the opposite: flowing from the sample back to noise, and integrating the divergence of the velocity along the path. When doing this under two models' (guided and unguided) velocity fields, the difference in log probabilities gives the odds ratio. 
    }
    \label{fig:pf-ode-schematic}
\end{figure}
The PF-ODE allows for the calculation of the (log) probabilities needed for the odds ratio. Essentially, for a given sample, we compute the likelihood of the sample twice: once using the guided model with the additional guidance term in the ODE velocity (see eq.~\ref{eq:cguidance}), and once without. The integration of the ODE will follow different trajectories $\mathbf{x}_t'$ (unguided) and  $\mathbf{x}_t$ (guided), along the flows  $\mathbf{v}_{\theta}'(\mathbf{x}_t', t)$ (unguided) and  $\mathbf{v}_{\theta}(\mathbf{x}_t, t)$ (guided). This results in the formula for the log odds ratio
\begin{align}
\log o(\mathbf{x}) &= \log(p_T(\mathbf{x}_T')) - \log(p_T(\mathbf{x}_T)) \nonumber \label{eq:divergenceint}\\
&\quad + \int_{0}^{T} \nabla \cdot \left[\mathbf{v}_{\theta}'(\mathbf{x}(t), t, A) - \mathbf{v}_{\theta}(\mathbf{x}(t), t)\right] \, dt 
\\
&= \underbrace{\log(p_T(\mathbf{x}_T')) - \log(p_T(\mathbf{x}_T))}_{\text{probability of latent under Gaussian prior }} \\
&\quad + \int_{0}^{T} \underbrace{\nabla \cdot \left( \frac{\mathbf{x'}- \mathcal{D}_{\theta}(\mathbf{x'}, \sigma)}{\sigma} -\frac{\mathbf{x}- \mathcal{D}_{\theta}(\mathbf{x}, \sigma)}{\sigma} \right)}_{\text{divergence of denoiser terms along different trajectories }} \, dt 
\label{eq:div1}\\ 
&\quad - \int_{0}^{T} \underbrace{\nabla\cdot \gamma \nabla_{\mathbf{x}} \mathcal{L}_{BCE}(\mathcal{C}(\mathbf{x}), y)}_{\text{divergence of guidance term } } \, dt \label{eq:div2}
\end{align}
For future reference, we will refer to the three difference terms above as 
\begin{equation}\label{eq:deltas}
    \log o(\mathbf{x}) = \Delta_{latent} + \Delta_{denoiser} + \Delta_{guidance}.
\end{equation}

Figure~\ref{fig:pf-ode-schematic} shows a schematic diagram of the trajectories starting from a sample $\mathbf{x}_0$ and flowing along the guided or unguided flow velocities towards the distribution of the Gaussian prior, where they end with different latents. It also shows the direction of the flow velocity components from the denoiser and the guidance term. 
The data sample $\mathbf{x}$ can either come from the training data distribution, or be a previously generated sample. The latter is the case that we are interested in here, where we have produced a guided sample that we want to find the odds ratio for. 

\subsubsection{Implementation of odds ratio calculation}\label{sec:implementation}

The PF-ODE trajectory is integrated using first-order Euler steps. Given $\mathbf{x}_i$ at noise level $\sigma_i$, the next state is:
\begin{equation}
\mathbf{x}_{i+1} = \mathbf{x}_i + (\sigma_{i+1} - \sigma_i)\, \mathbf{v}_{\theta}(\mathbf{x}_i, \sigma_i)
\end{equation}
The divergence terms eqs.~\ref{eq:div1}, \ref{eq:div2} are approximated using the Skilling-Hutchinson estimator from eq.~\ref{eq:trace_estimator} with the noise $\boldsymbol{\epsilon}_k \sim \mathcal{N}(\mathbf{0}, \mathbf{I})$ and the Jacobian-vector products $\nabla {\mathbf{v}}_{\theta} \ \boldsymbol{\epsilon}$ are computed using automatic differentiation. We average this estimator over $k \in [0,1,2]$ draws from the normal distribution, trading off variance of the estimator against compute.

The integral in eq.~\ref{eq:divergenceint} is approximated using the trapezoid rule:
\begin{equation}
\int_{\sigma_{\min}}^{\sigma_{\max}}
\nabla \cdot \mathbf{v}_{\theta}(\mathbf{x}(\sigma), \sigma)\, d\sigma
\approx
\sum_{i=0}^{M-1}
\frac{\hat{d}_i + \hat{d}_{i+1}}{2}
(\sigma_{i+1} - \sigma_i)
\end{equation}
where $\hat{d}_i = \nabla \cdot \mathbf{v}_{\theta}(\mathbf{x}_i, \sigma_i)$ and we discretize the ODE trajectory into $M$ steps. The $M$ steps consist of 36 steps spaced according to the sampler in EDM, plus 20 additional equally spaced ODE time steps inside the guidance window $\sigma \in [15, 20]$.

When calculating the odds ratio, we could, in theory, calculate the probability under the guided model already during the sampling (backward ODE) and only the unguided probability using the forward ODE, overall flowing from prior to data to prior. However, since the discretization of the ODE can introduce errors and a dependence on direction, we here first generate the guided sample $\mathbf{x}$, and then calculate the two probabilities $p_{\text{unguided}}(\mathbf{x}), p_{\text{guided}}(\mathbf{x})$  using the forward ODE by flowing from data to prior using the same sampling steps (here, we can therefore not use a black box ODE solver with adaptive step size).

\subsection{ERA5 sample likelihood}\label{sec:methods-cc}
\subsubsection{Single-channel Antarctic student model}
A more standard calculation is that of the likelihood $\mathbb{E}_{\mathbf{x} \sim p_{data}}[\log p_\theta(\mathbf{x})]$. We here apply the probability calculation to samples of ERA5 data. We want to test the dependence of the sample probability on extreme events as well as on different SST input conditioning of the model. However, to separate this signal from the noise, we need to restrict the sample to the extreme event, using only the surface temperature variable and only in a single region. This leaves us with a relatively small fraction of the total pixels of the full atmospheric state. 
We want to calculate the terms of eq.~\ref{eq:pfode} only on these pixels. For this we retrain a student model on cBottle output for the single variable temperature at surface (tas). For the region of interest we choose the Antarctic, allowing us to study temperature anomalies there. The student model `sees'  only the tas values in the Antarctic south of $65^{\circ}$S. As training data, we reuse a cBottle inference over the AMIP period, i.e. three-hourly states conditioned on SSTs between 1940 and 2017. This approach of training on synthetic data is inspired by the work of  \citep{martin2025longrangedistillationdistilling10000}, who use a similar method to make long lead time forecasts. The inputs to the model are the same, and the architecture is unchanged except for a reduction in model channels to 64. For convenience, we still use the global HEALPix grid in space, masking the area outside the Antarctic with zeros. Training takes about a fifth of the compute as for the full cBottle model. Note that we could have also trained on the original ERA5 data, but that the synthetic inference data with its many more independent samples is helpful to avoid the model's tendency to overfit to large scales. 

Figure~\ref{fig:sst-pert} shows a comparison of the responses of our student model (a) and the cBottle teacher (b) to an imposed uniform SST perturbation of +2K (applied only to the sea pixels, land pixels retain the constant `masking' value used in training of 290K). The plot of the climatological anomaly over the AMIP times shows an unphysical land cooling over Antarctica. This is consistent with results from cBottle runs reported in the AI model intercomparison project (AIMIP, forthcoming). Meanwhile, the student model warms everywhere, albeit only to a small degree. 

\subsubsection{Probability density calculation}
For probability density calculations, we follow \cite{song2020score} in using the SciPy \texttt{scipy.integrate.solve\_ivp} black-box ODE solver to calculate the trajectory, which uses a 5th order adaptive Runge-Kutta scheme (\texttt{rk45}), performing six function evaluations per integration step. This more computationally expensive choice of solver is possible here because the function evaluations are cheaper: there is no guidance, and therefore no double-differentiation through the classifier network. 



\section{Results}\label{sec:results}
\subsection{Guidance in a small denoising interval}\label{sec:results-window}
\begin{figure}
    \centering
    \includegraphics[width=0.75\linewidth]{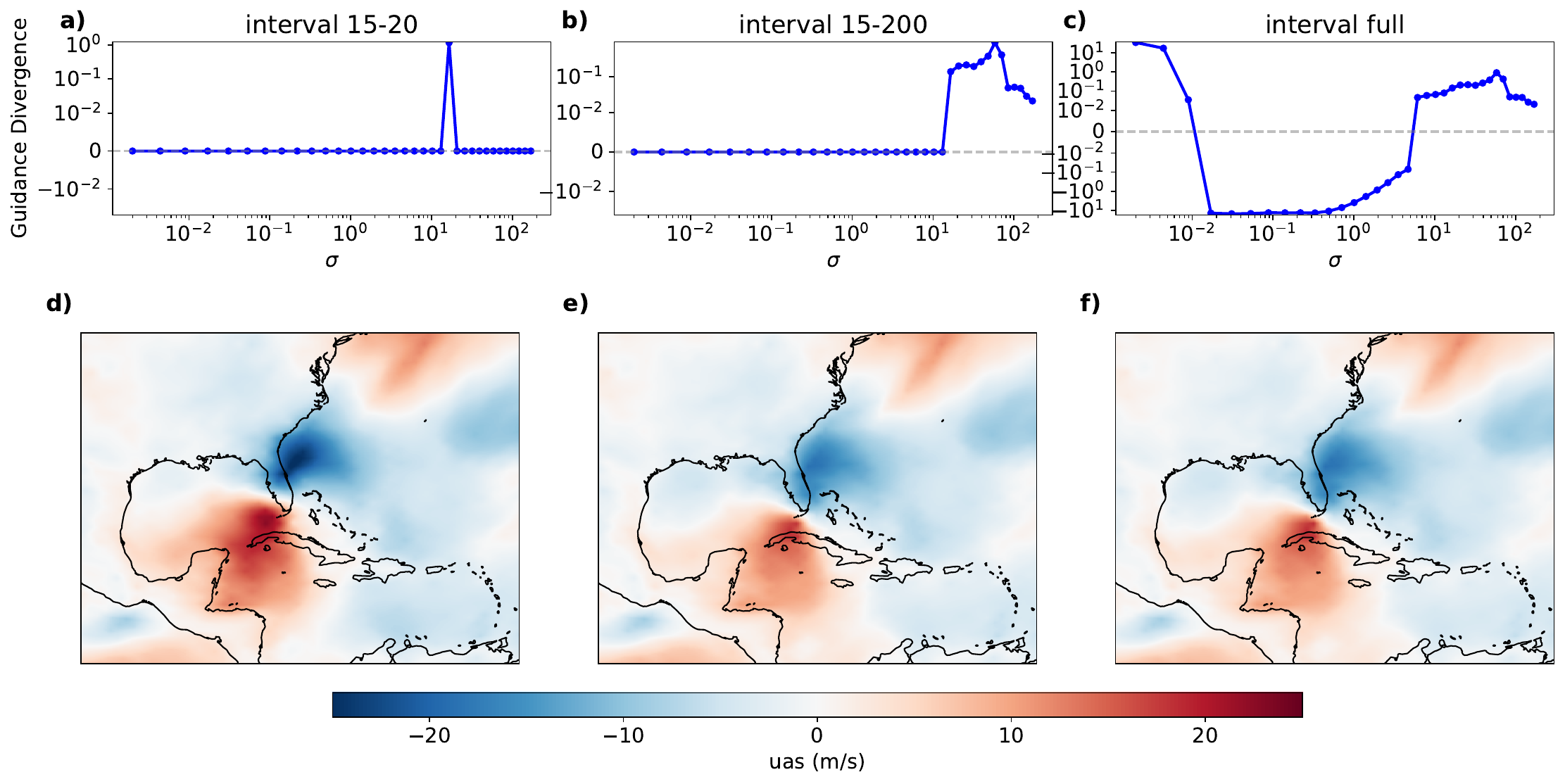}
    \caption{Impact of guidance interval on guidance divergence and samples (EMA MoE denoiser and EMA TC classifier), for guidance towards a TC in Miami, Florida. Columns show guidance restricted to the given in each column's title. The top row plots the instantaneous divergence of the guidance term at each noise level, which integrated over the denoising time gives $-\Delta_{guidance}$. The bottom row shows the resulting zonal wind in the region where guidance is applied.}
    \label{fig:guide-interv}
\end{figure}
Before moving to calculating the probability of TC states, we present a number of results on how to implement guidance. In order to oversample TCs in the data distribution, we are free to adjust the guidance according to what method yields the most stable calculations of the PF ODE. Note that we simplify the guidance in comparison to \cite{cbottle}, by replacing the adaptive guidance scale with a constant (see section~\ref{sec:guidance-method}). Separately, we can choose the denoising interval during which to apply guidance, with the default being to guide along the whole trajectory. 

However, we find that applying guidance along the entire trajectory leads to implausible values of the guidance divergence term. This term, expressing whether the probability density is increasing or decreasing, is expected to be positive (intuitively, when denoising, guidance `converges' trajectories to become more similar). In Fig.~\ref{fig:guide-interv}, top row, the guidance divergence is plotted as a function of the noise level (equivalent to denoising time) for different choices of the guidance interval. The state resulting from this guidance is shown in the bottom row. Panels c), f) show the default full interval guidance, with values of the divergence being negative below a noise level of around two. This is unexpected: We expect the overall log odds ratio to be negative for a TC sample (and in particular one obtained under guidance; see the sign convention in section~\ref{sec:meth-is}), meaning it is more likely that such a state was obtained with guidance than without. Negative contributions in some noise levels seem to indicate that the guidance leads to more diverse samples there, rather than focusing samples on a high-TC-probability subset. 

Fortunately, we are free to exclude this region by setting the guidance to zero below $\sigma=15$ (panels b), e)). This results in positive guidance divergence values throughout, \emph{without noticeably changing the structure of the guided sample}. This reduction of the guidance window can be made even more drastic, by restricting to the interval $\sigma \in[15, 20]$ (panels a), d)). The ODE being discretized during sampling, this corresponds to a single step of guidance, which panel d) shows to be enough to result in a guided sample. 

The finding that only a comparatively small interval is relevant for the generation of TCs is consistent with our expectation: \cite{cbottle} found that high noise levels were needed in training and inference of cBottle to `cover' the signal of the seasonal and diurnal cycle. Conversely, this means that these signals are generated early in the denoising, at high noise levels ($\sigma \approx 200$). Smaller signals (both in amplitude and spatial scale) such as TCs will then be generated at intermediate noise levels, with the smallest scale details being produced at very low noise. Different semantic features correspond to different parts of the denoising trajectory. This is in agreement with the results of \cite{galashov2025learn}, who show that in classifier-free guidance on ImageNet, there are diffusion time intervals where guidance is more important than in others, and that the sensitivity depends on the condition that is being enforced. 

From a computational perspective, guiding only on a small interval is advantageous when the guidance is expensive (here, it requires a differentiation through the network, and a second one for the calculation of the divergence). Here, we go forward with the choice of the interval $[15, 20]$. 

\subsection{Large variances of sample-wise odds ratios}
\begin{figure}
    \centering
    \includegraphics[width=1\linewidth]{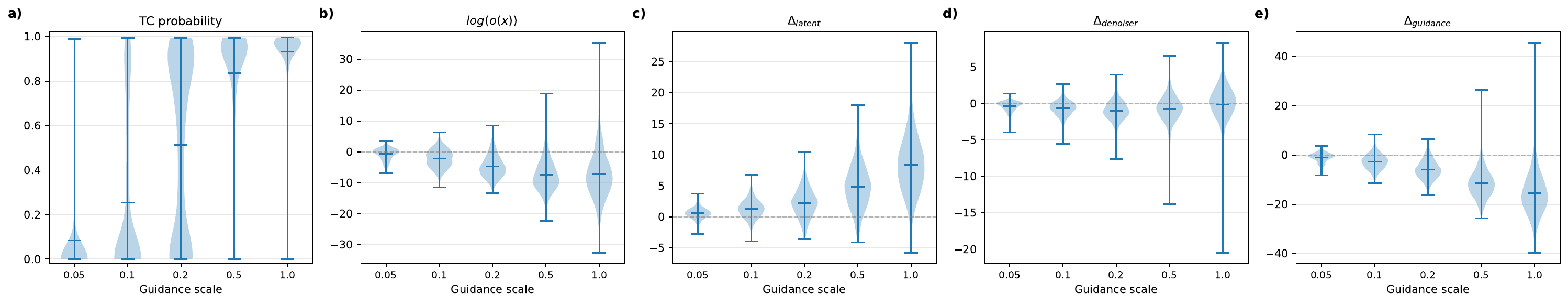}
    \caption{Distribution of TC probabilities and log ORs from different guidance strengths (EMA MoE denoiser and EMA classifier, guidance restricted to $\sigma\in[15,20]$), motivating the use of small guidance. Shown are distributions of a) the classifier evaluated on the final sample, i.e. the detection probability of a TC at the location where guidance was applied, b) the log odds ratio $\log o(x)$ of the final sample, which is the sum of c) the difference of final log probabilities $\Delta_{latent}$ d) the (integrated) difference of score divergences $\Delta_{denoiser}$, and e) the (integrated) divergence of the guidance term $\Delta_{guidance}$ from eq.~\ref{eq:deltas}. Guidance scale on the x-axis is given as a fraction of the full guidance of $\gamma=20\cdot64$. Inference is performed over 704 samples with boundary conditions of September 2004-2009, 6-hourly, guidance towards a TC off the coast of Florida as in Fig.~\ref{fig:guide-interv}. }
    \label{fig:lower-guidance-scale}
\end{figure}
In order to reduce the variance of the importance sampling estimator from eq.~\ref{eq:is}, we find it to be important to tune the guidance strength. We find that for guidance strengths where practically each generated sample shows a TC ($\gamma = 64 \cdot 20$), the calculated odds ratio differs a lot from one sample to the next. Figure~\ref{fig:lower-guidance-scale} shows distributions of the classifier output at the guidance location when evaluated on the final sample, as well as the corresponding odds ratios and its components, for different guidance scales. Panel a) illustrates how the TC probability in the sample strongly depends on the guidance strength, as is expected: Stronger guidance leads to more samples with high TC probability. Crucially, though, even in the unguided inference ($\gamma=0$), there are still some TCs present in the location of interest. 

The odds ratios in panel b) are consistent with the change in TC probability: They express how much less likely the guided samples would be under the unguided model than under the guided one. Hence, the larger the guidance and the smaller the overlap between the guided and unguided distribution, the more negative the odds ratio. However, it also shows a heavy tail towards very positive values. This is surprising, as guided samples should only rarely be more likely under the unguided model. The odds ratios show large variance in the strong guidance cases. These long tails are suppressed as guidance strength is reduced and overlap between the distributions increases. 

Decomposing the odds ratios into its constituent terms (panels c)-e)) shows that the pattern of increased variance for stronger guidance is similar in all terms, with the strongest outliers coming from the difference in final log probability of the noisy sample and the guidance term. Signs of each term are as expected given the convention of section~\ref{sec:meth-is}: guidance leads to a lower final log probability of the latent under the Gaussian prior ($\Delta_{latent}<0$), a more negative score divergence ($\Delta_{denoiser}<0$), and a positive guidance divergence ($-\Delta_{guidance}>0$, as plotted).

We note that discretization errors and imperfectly trained networks for the denoiser and classifier may contribute to the spread in odds ratio estimates and go forward with a reduced guidance scale of $\gamma=0.1\cdot64\cdot20$.
\subsection{Importance-sampled extreme event likelihoods}
\begin{figure}
    \centering
    \includegraphics[width=0.9\linewidth]{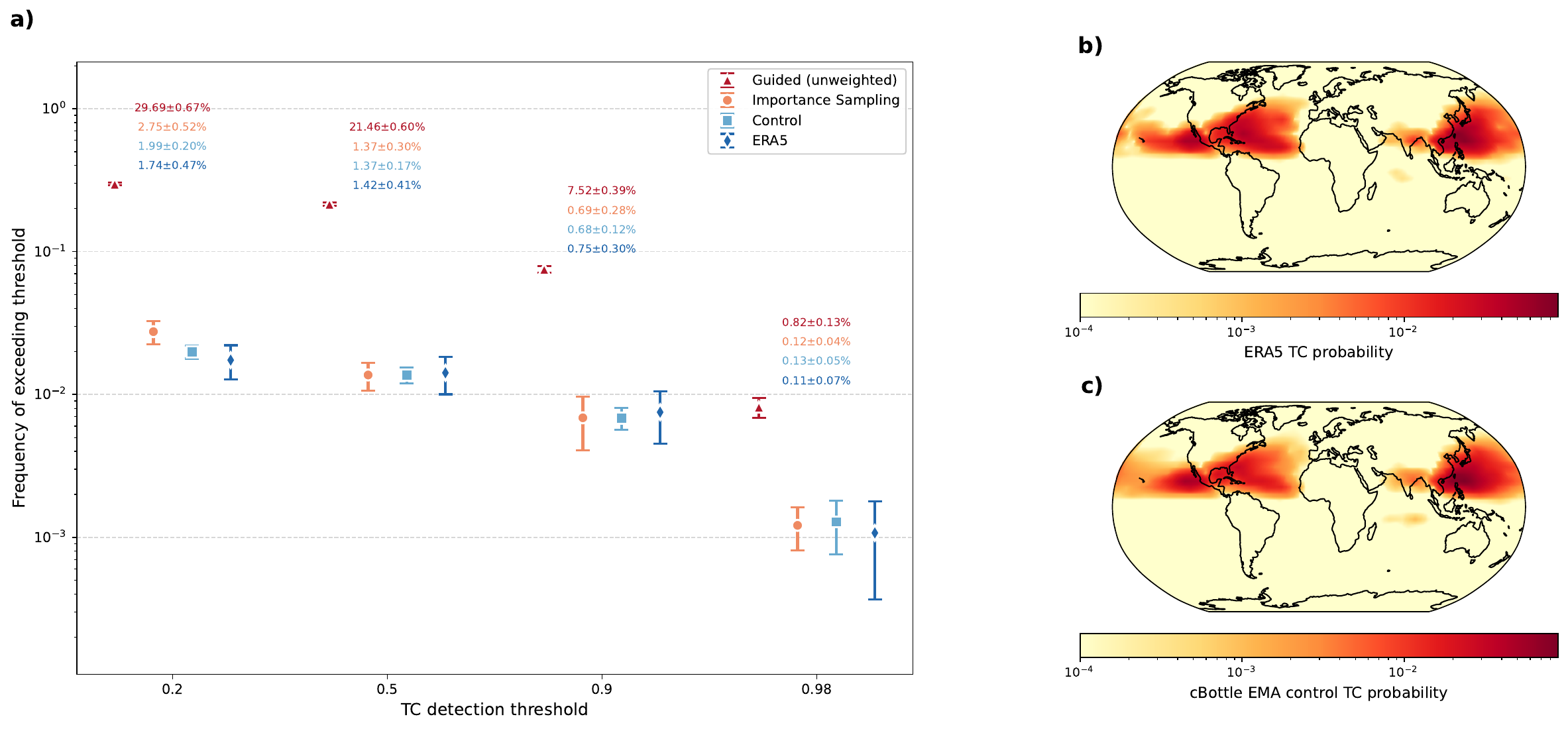}
    \caption{Comparison of ERA5 and cBottle TC climatology and importance-sampled TC likelihoods (EMA MoE denoiser and EMA classifier). Panel a) shows the estimates of the exceedance frequency of detection thresholds 0.2, 0.5, 0.9, and 0.98. We are counting a detection when the TC classifier returns a probability above the threshold in the location of interest. The frequency of detection is computed from different data sets: The guided cBottle model (oversampled), the importance sampling estimate (guided, but downweighted by odds ratios), the control cBottle (unguided), and the historic record from ERA5. The frequencies are for occurrence in a specific location, chosen here to be the HPX3 pixel including Miami, Florida. We evaluate on the September months of the years 1980-2018 inclusive (4677 samples). Panel b) shows the climatology of September TC detection probabilities (average) over the global ERA5 record of the same period. c) shows the same for the control inference. 
    }
    \label{fig:is-tc}
\end{figure}
We now turn to using the calculated odds ratios for importance sampling. We show that, even though we need to adjust the guidance as described above, we can retrieve accurate probabilistic information from guided samples. We run two large inferences over 40 years of September SSTs as boundary conditions, yielding 4797 samples from the guided and unguided model respectively. We evaluate the samples with the classifier for the detection probability of a TC in the location of interest. We can choose a threshold for detection of a TC, e.g. 50\%. We now obtain two estimates for the frequency with which cBottle produces states with a TC detection: One from counting the detections in the unguided inference as in eq.~\ref{eq:mc}, and one from counting the detections in the guided inference and multiplying each count with the sample odds ratio, as in eq.~\ref{eq:is}. These estimates, together with the unweighted frequency of exceeding the detection threshold in the guided inference, are shown in Fig.~\ref{fig:is-tc} a), for different choices of the threshold. Comparing the unweighted guided estimate (red) with the unguided control (light blue), we see that the guidance has increased the TC frequency by about one order of magnitude. 

Comparing the estimate for TC frequency from the control inference (light blue) with the importance sampling estimate (orange) shows that the calculated odds ratios successfully downweight the frequency of the guided states. The frequency is downweighted to agree with that from the control, within two standard errors for the lowest, and one standard error for all higher (i.e. more extreme) detection thresholds. Crucially, the relative uncertainty of the importance sampling estimate increases more slowly when moving to more extreme events than that of the control, because the control's uncertainty is dominated by counting noise on very few detections (as few as 6 out of 4677 samples at the highest threshold shown, 0.98) while the importance-sampling estimate spreads that uncertainty across many more classifier-positive guided samples (38 at the same threshold). Therefore, at the highest threshold of 0.98, the importance sampling estimate reaches a lower relative uncertainty than the control (relative uncertainty $34\%$ IS vs.\ $41\%$ control), even though its point estimate agrees with both the control and ERA5 within one standard error. This demonstrates how oversampling the distribution with guidance for extremes can lead to more confident estimates of extreme event likelihoods, though the improvement should be interpreted with the small raw counts in mind. We further stratify this comparison by physical TC intensity, rather than classifier confidence, in section~\ref{sec:tier3-intensity}, showing that using a thresholding that is correlated with, but not exactly the guidance variable, introduces additional uncertainty in the estimate.

Accurate extreme event likelihoods depend not only on sampling, but also crucially on the accuracy of the underlying model. Here, we compare our agreeing estimates of TC probabilities with the historic record from ERA5 data. Applying the classifier to the ERA5 data of the same time stamps that the SSTs were sampled from yields the estimate from the historic record shown Fig.~\ref{fig:is-tc}, with panels b) and c) comparing the spatial pattern of the time-averaged classifier output, and the dark blue markers in panel a) showing the TC probability estimates in the location of interest. ERA5 frequency uncertainties are block-bootstrapped. Spatially, the cBottle inference matches well the historic record, and after swapping raw model weights with EMA weights, frequencies agree between ERA5, control, and IS estimate at all thresholds. For comparison, the same Figure for the raw model weights is reproduced in section~\ref{sec:app-raw-weights}.

Unfortunately, while importance sampling reduces the sampling error for a fixed sample size, the odds-ratio is expensive to compute and in practice it is cheaper to simply draw more unguided samples. Taking an unguided inference of cBottle as the reference, which takes about 15 sec on an NVIDIA H100 GPU, we incur a slowdown for guidance by 2x, a slowdown of 1.5x for the extra steps needed to accurately calculate integrals along the sampling, and a slowdown of 11x for the two forward ODE integrations with calculation of the odds ratio (evaluating the divergences of the guided and unguided score). This gives an overall slowdown of 33x for a single guided sample with odds ratio for this proof-of-concept implementation. We discuss ways of reducing this cost in the conclusions. 

We report the ablation that justifies these sampler settings, an ablation of whether the $[15,20]$ guidance window is necessary for an accurate estimate, and the re-stratification of this comparison by physical TC intensity rather than classifier confidence, in Appendix sections~\ref{sec:tier1-cost}, \ref{sec:tier2-windows}, and \ref{sec:tier3-intensity} respectively.

\subsection{Outlook: ERA5 sample likelihoods and climate change attribution}
We want to briefly discuss another application of the probability calculated using a trained diffusion model: calculating the likelihoods of samples from data with a view to extreme events and climate change attribution. We flag upfront that the result of this section is largely exploratory and negative: as we show below, the diffusion-model log-likelihood does \emph{not}, in this application, track the extremeness of the sampled temperature anomaly. The weak-but-correct-sign SST sensitivity in panel d) below is still informative about what these likelihoods can and cannot resolve, but does not demonstrate a working extreme-event-attribution tool.
\begin{figure}[h]
    \centering
    \includegraphics[width=0.9\linewidth]{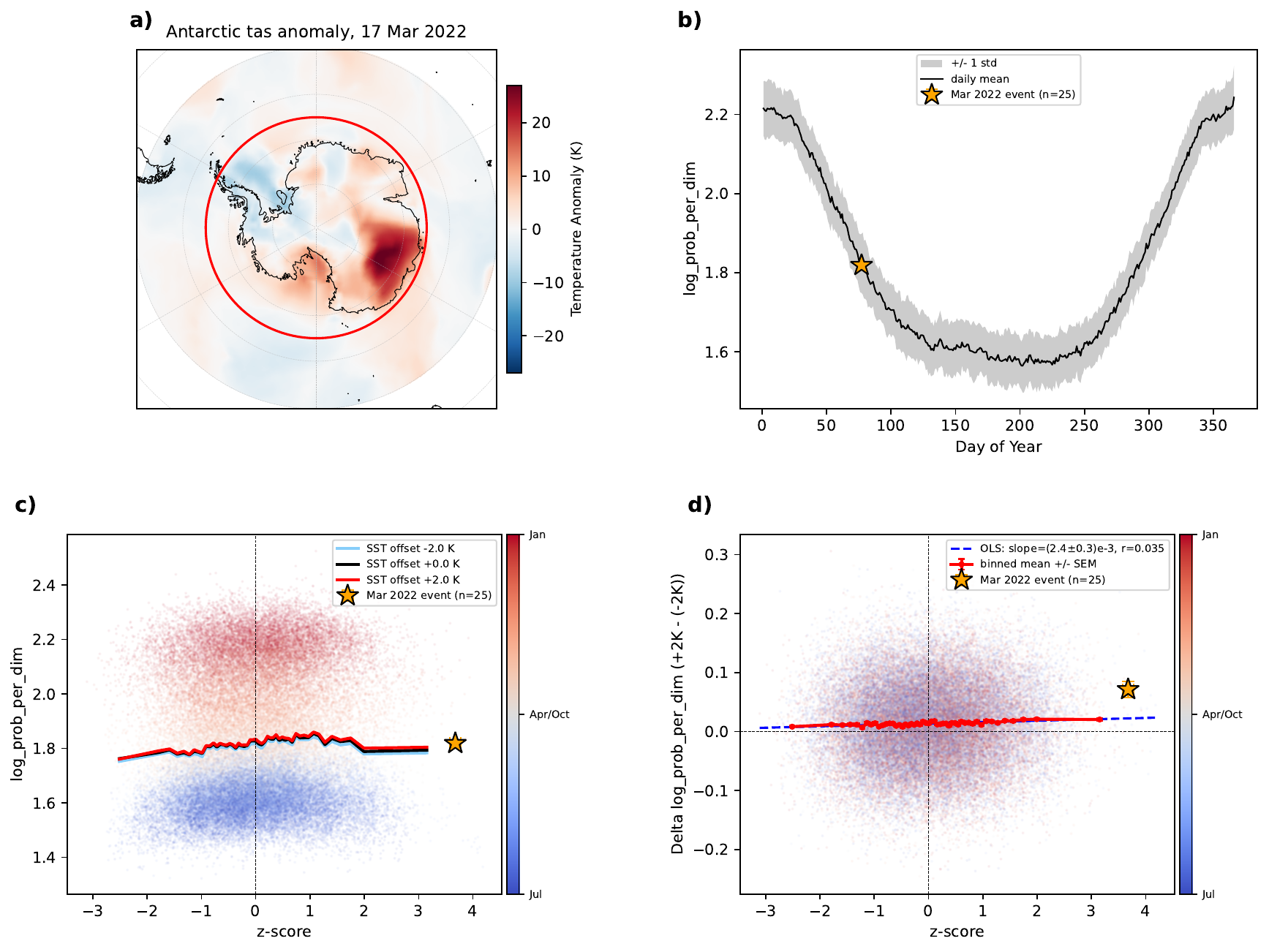}
    \caption{Calculating likelihoods for ERA5 reanalysis samples, here for heat events in the Antarctic. Panel a) shows a visualization of the heatwave surface temperature anomaly, on 17 March 2022 at 00:00 UTC from ERA5. The red circle shows the unmasked region of interest. Panel b) shows the seasonal cycle of sample probabilities over the entire train set, with the shaded area showing the standard deviation. Panel c) plots the sample probability as a function of the z-score of the average tas anomaly over the region, as well as binned means of the probability for the data as-is, and with SST inputs to the model perturbed by ±2K. Panel d) shows the difference between the +2K and the -2K perturbation, as well as the binned mean and a linear regression (difference between the red and blue line in panel c)).}
    \label{fig:cc-antarctic}
\end{figure}
As an example, we take the Antarctic heatwave of March 2022, which, peaking at −10.1°C, presented a 39°C departure from climatology, making it in some sense the largest heatwave on record \citep{blanchard2023largest}. Note, that this extreme occurred in austral fall; the same temperature maximum would have been less extreme in the summer. It was skillfully forecast, and linked to a large-scale weather pattern importing large warm air masses from Australia. We show the temperature anomaly from ERA5 in Fig.~\ref{fig:cc-antarctic} a).  

Instead of calculating the probability of a guided generated sample under the guided and the unguided model as in the previous section, we here calculate the probability of a (real, ERA5) data sample under the (unguided) model. We study how the distribution of probabilities changes depending on the sample. 

The temperature deviation from climatology is not systematically linked to the log probability. Indeed, panel (b) shows that the extreme March heat wave is assigned typical log probability values for that time of year. Fig.~\ref{fig:cc-antarctic}c), show the sample log probability against the sample's z-score
\begin{equation}
    z_{ij} = \frac{T_{i,j} - T_{i, \ clim}}{std(T_{i})_{clim}}
\end{equation}
with the region-average temperature at surface $T_i$ on day $i$ of year $j$, the climatological average value $ T_{i, \ clim}$, and the standard deviation of $T_i$ across the climatology years. The scatter plot shows the distribution of summer (higher probability) and winter (lower probability) points, but the binned means (solid lines) do not show smaller probabilities for more extreme z-scores, as we may have expected. The binned average is plotted for the model with the SST conditions of the sample (SST offset 0K), as well as for the case of perturbed SST conditions by ±2K. The difference in the probability as a function of z score is shown in d), with a moderate increase in the gap for heat events—the model correctly assigns higher probability to the +2K than the -2K condition for a heat wave such as the March 2022 event. Smaller, but still positive differences in probability are also assigned for cold extremes. 

Fig.~\ref{fig:cc-antarctic}b) shows that there is a pronounced seasonal cycle in the sample probability, with the highest probabilities assigned to samples in austral summer, and the lowest in winter. The negative probability can be viewed as a measure of how much information needs to be stored for this particular sample when compressing the data \cite{shannon1948mathematical}. A sample with high probability is more easily compressed. Therefore, we expect higher probabilities when natural variability is small. In winter, average temperatures in the Antarctic are more variable than in the summer (standard deviations of the anomaly with respect to climatology of 2K in the summer vs. 1K in winter), possibly linked to sea ice variability and the Southern Hemisphere storm tracks. 

In sum, while the diffusion estimated probability of the extremes shows some intriguing dependence on SST the assigned log probability does not effectively correlate with the presence of extreme events.

\conclusions\label{sec:conclusions}  

In this study, we have shown that we can retrieve probabilistic information from a diffusion model guided to oversample extreme events, using the cBottle model guided to Tropical Cyclones. This allows us to sample more extreme events, while obtaining an estimate of how likely these events are. We can calculate likelihoods of TCs occurring in a given location using importance sampling, which match a Monte Carlo estimate from cBottle. To our knowledge, this calculation of odds ratio as the ratio of sample probability under a guided and unguided model, has not been performed previously for any guided diffusion model. The accurate odds ratios increase our confidence in extreme event guidance being consistent with the data distribution. We find that it is sufficient to apply guidance at a single sampling step corresponding to the magnitude of the extreme event. 

The calculated odds ratios, describing how much more likely a TC sample is under the guided than under the unguided model, should be accurate for individual samples. However, we find that, with our method, there can be large variance in the odds ratios between similar samples. This is exacerbated for stronger guidance, so importance sampling is only successful for weaker guidance and sampling repeatedly. Nevertheless, we show that for extremes, the uncertainty of the importance sampling estimate can be smaller than that of the unguided Monte Carlo estimate. The mechanism for this is by replacing sampling uncertainty (few extreme events) with odds ratio uncertainty (high variance of weights). The former grows faster than the latter with increasingly extreme events, resulting in a better estimate for the rarest events. We show that for importance sampling, it is beneficial to guide on the variable of interest—guiding on TC probability but evaluating on wind speeds introduces additional uncertainty.

Currently, these estimates are more computationally expensive than increasing the number of samples for the Monte Carlo estimate: To narrow the uncertainty by one third, we incurred a slowdown of 33x compared to the Monte Carlo estimate. This can be compared with the 2.25x more inferences Monte Carlo sampling would require to reduce the uncertainty by the same amount. The comparison becomes more favorable for more extreme events, and so could flip in favor of importance sampling for even rarer extremes than presented here. Also, these initial performance estimates should be interpreted with nuance appropriate to a research prototype. Most of the computational cost for these inferences comes from the many sampling steps, so that distillation of the probability calculation \citep{distillation1, distillation2} seems promising. In a best case scenario, this would lead to a single step evaluation of the probability, resulting in an overall slowdown of only 2x (one calculation for the guided and one for the unguided model, respectively). We therefore expect the net effect of these methods in the long term to be more cost-effective sampling of extremes. 

Besides the importance sampling of extreme events, we demonstrate a further application of probabilities calculated from a trained diffusion model: For ERA5 samples, we calculate the likelihood under the model and obtain a consistent, if too-weak, response of probabilities to SST conditioning. We note that this method of computing probabilities could be used to evaluate the KL-divergence between two data distributions $p$ and $q$, e.g. the observed data $p$ and outputs of an AI weather model $q$. This would require an additional diffusion model trained on samples from $q$ to evaluate the term $\mathbb{E}_{\mathbf{x} \sim p}[\log q(\mathbf{x})]$ from eq.~\ref{eq:kl-div}.


There is a number of possible avenues for further research: For the likelihoods of extreme events, in particular TCs, it may be beneficial to change the way guidance is implemented. The current classifier guidance requires differentiating through the classifier for calculating the guidance. This means that for estimating probabilities, where the divergence of the guidance is required, we need to differentiate through the network twice. This is computationally expensive and a possible source of errors. Classifier-free guidance, for instance, would not require a first differentiation for the guidance term (but would require a retrained conditional denoiser). This could be combined with different approaches sampling and computing the odds ratio, such as the Radon-Nikodym estimator for Sequential Monte Carlo sampling \citep{rne}.

Training additional models could be useful for some applications. For example, to accurately compute climate counterfactuals, it would be necessary to train on climate simulations with pre-industrial and potential future SSTs. In order to mask out regions, spatial dropout could be included in training. Improvements in the model, in particular in its representation of TC strengths, would improve estimates of extreme event likelihoods independently of the guidance and sampling methods studied in this work. 

We hope that the advances around the unique applications of diffusion model climate emulators presented here will stimulate new research into these methods, making them a helpful, reliable, and widely accessible tool for climate research.


\codedataavailability{ERA5 data is freely available from the Copernicus climate data store. The trained cBottle model checkpoints (both EMA- and non-EMA versions) are available from huggingface under \url{https://huggingface.co/nvidia/cbottle}. Training and inference code is available under \url{https://github.com/NVlabs/cBottle/tree/main}, and inference, analysis, and plotting scripts specific to this project will be made available there at the time of acceptance of the manuscript.} 

\newpage
\appendix
\section{Extended results}    
\subsection{Bound on individual sample probability density accuracy}\label{sec:app-pinsker}
Here, we expand the step from \ref{eq:sample-bound}, showing that
\begin{equation}
    2\,\mathbb{E}_{p_{data}}\left[\left(\log \frac{p_\theta(\mathbf{x})}{p_{data}(\mathbf{x})}\right)_+\right]
    \le \sqrt{2\,D_{KL}(p_{data}\parallel p_\theta)},
\end{equation}
where $(\cdot)_+=\max(\cdot,0)$.   

First, for any $t>0$, $\log t \le t-1$ implies $(\log t)_+ \le (t-1)_+$.
Setting $t=p_\theta(\mathbf{x})/p_{data}(\mathbf{x})$ and taking expectations under $p_{data}$,
\begin{equation}
    2\,\mathbb{E}_{p_{data}}\!\left[\left(\log \frac{p_\theta(\mathbf{x})}{p_{data}(\mathbf{x})}\right)_+\right]
    \;\le\;
    2\,\mathbb{E}_{p_{data}}\!\left[\left(\frac{p_\theta(\mathbf{x})}{p_{data}(\mathbf{x})}-1\right)_+\right].
\end{equation}
 Second, writing out the expectation as an integral and pulling $p_{data}$ inside,
\begin{equation}
    2\,\mathbb{E}_{p_{data}}\!\left[\left(\frac{p_\theta(\mathbf{x})}{p_{data}(\mathbf{x})}-1\right)_+\right]
    = 2\int \bigl(p_\theta(\mathbf{x})-p_{data}(\mathbf{x})\bigr)_+ \,d\mathbf{x}
    = 2\,TV(p_{data},\,p_\theta),
\end{equation}
using $TV(p,q)=\int(p(\mathbf{x})-q(\mathbf{x}))_+\,d\mathbf{x}$, which holds because
$\int p=\int q=1$ forces the positive and negative parts of $p-q$ to have equal mass.

 Third, Pinsker's inequality states $TV(p,q)\le\sqrt{D_{KL}(p\parallel q)/2}$, so
\begin{equation}
    2\,TV(p_{data},p_\theta)
    \;\le\;
    2\sqrt{\frac{D_{KL}(p_{data}\parallel p_\theta)}{2}}
    = \sqrt{2\,D_{KL}(p_{data}\parallel p_\theta)}.
\end{equation}
Chaining the three inequalities gives the stated bound.
\subsection{Estimator uncertainty and computational cost}\label{sec:tier1-cost}
The odds-ratio estimate in section~\ref{sec:implementation} depends on several numerical choices: the number of base sampling steps, the number of extra steps densifying the guidance window, and the number of Hutchinson probes $k$ used to estimate the divergence. We ablate these choices at fixed guided samples (rescoring 500 saved samples rather than regenerating them, so that the ablation isolates numerical error from sampling variability), varying the base step count (18, 36), the number of extra steps in $[15,20]$ (0, 5, 25), and $k$ (1, 3, 6), plus one comparison point using Heun rather than Euler integration at the paper's setting (36 base $+$ 25 extra, $k=3$). We also measure wall-clock GPU-seconds per sample directly.

\begin{figure}
    \centering
    \includegraphics[width=0.95\linewidth]{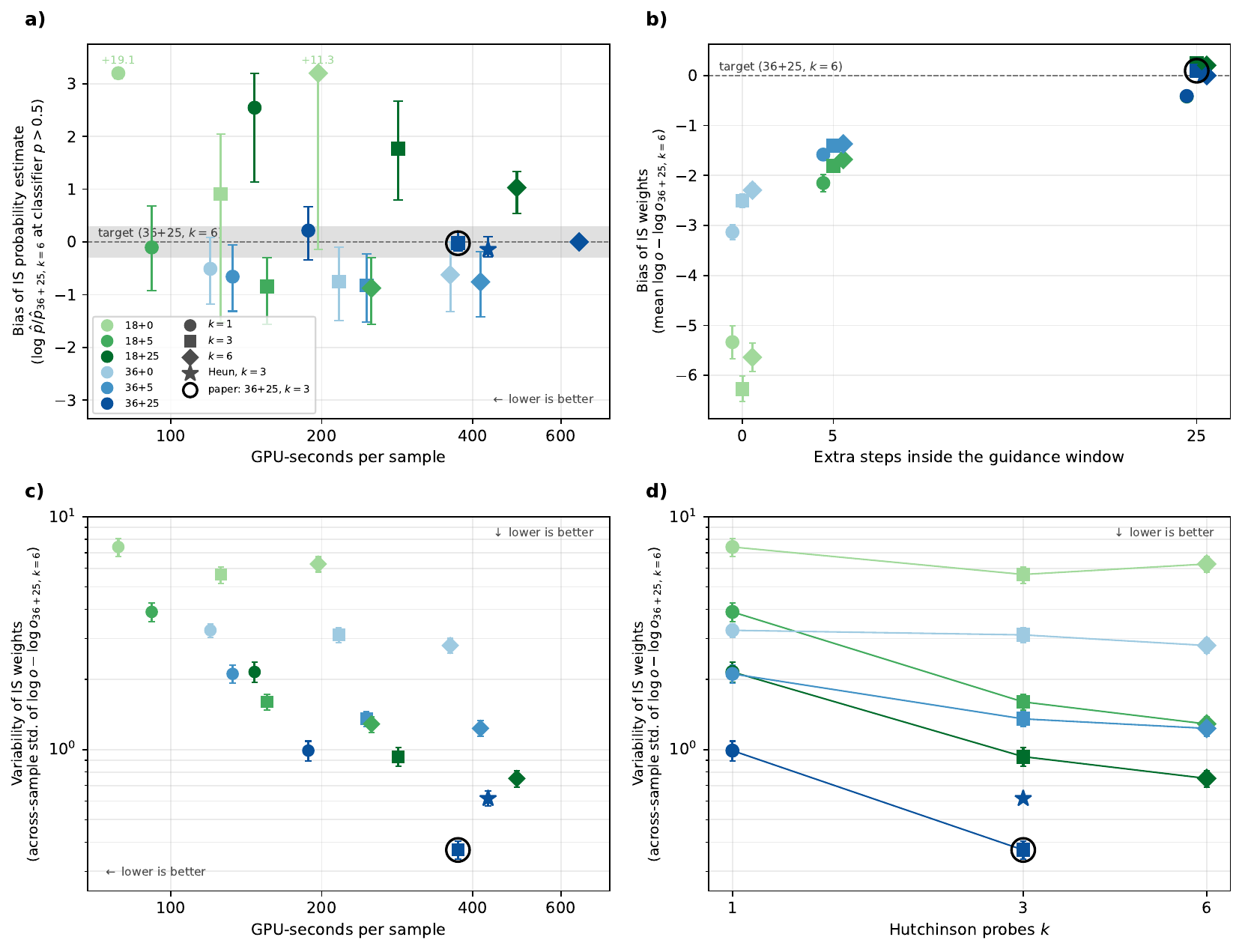}
    \caption{Cost-normalized accuracy and uncertainty of the odds-ratio estimator across 19 sampler settings (500 rescored samples each), relative to the densest setting (36 base steps, 25 extra steps, $k=6$). a) $\log(\hat{p}/\hat{p}_{\mathrm{ref}})$ at classifier $p>0.5$ versus GPU-seconds. b) Mean $\Delta\log o$ versus number of extra steps in $[15,20]$. c) Across-sample standard deviation of $\Delta\log o$ versus GPU-seconds on NVIDIA H100 GPUs. d) Same spread versus number of Hutchinson probes $k$. The paper setting (36 base $+$ 25 extra steps, $k=3$) is ringed in each panel. Error bars: bootstrap 95\% CI (a, c, d); analytic SEM (b).}
    \label{fig:tier1-sampler}
\end{figure}

Figure~\ref{fig:tier1-sampler} shows the bias of the IS estimate (a) and the bias (b) and variability (c, d) of the IS weights for the different settings of the ablation. All comparisons are with the highest-cost calculation, implicitly assuming that increasing integration steps and probes over the control-comparable IS estimate of the paper setting only further reduces bias and variance. Panel a) shows that the paper setting is the cheapest setting which allows for an unbiased, low-variance IS estimate. The other panels break this down further, with b) showing that densifying the guidance window with extra steps is required to avoid biasing the mean weight estimate. Panel c) shows the variability of the weights against the computational cost, justifying the doubling of base steps from 18 to 36. Panel d) is very similar, plotting the variability as a function of Hutchinson probes k (which computational cost depends on almost linearly). The panel shows that there are diminishing returns to increasing k above the paper setting of 3. 

Heun integration at the paper setting increases GPU-seconds by only about 17\% relative to Euler, because the Hutchinson trace is not re-evaluated on the Heun correction step, but does not materially change $\hat{p}$ or its spread at this step count.

\subsection{Is the guidance window necessary for an accurate estimate?}\label{sec:tier2-windows}
Section~\ref{sec:results-window} (Fig.~\ref{fig:guide-interv}) shows that restricting guidance to $\sigma\in[15,20]$ avoids implausible values of the guidance divergence at low noise levels, and we adopted this window throughout. Here we test directly whether this restriction is necessary for the accuracy of the resulting importance-sampling estimate, since the IS target $p_{\mathrm{cBottle}}(TC)$ does not depend on the guidance window used to propose samples: in principle a full-trajectory guidance should still recover the same $\hat{p}$.

We compare four windows over the same 704 September times used for Fig.~\ref{fig:lower-guidance-scale}: the control ($\gamma=0$), the paper's $[15,20]$ window at $\gamma=0.1\cdot20\cdot64$, guidance over the full trajectory $[0,200]$ at the same $\gamma=0.1\cdot20\cdot64$ (unmatched, mean classifier $p=0.91$), and guidance over the full trajectory at a reduced $\gamma=0.01\cdot20\cdot64$ chosen to match the $[15,20]$ window's mean classifier probability ($p\approx0.25$).

\begin{table}[h]
\centering
\caption{Guidance-window comparison at 704 samples: mean and spread of $\log o$, fraction of samples clipped at $\log o \ge 5$, and the resulting exceedance-frequency estimate at classifier $p>0.5$.}\label{tab:tier2}
\begin{tabular}{l|llll}
\tophline
window & mean $p$ & mean $\log o$ & std $\log o$ & $\hat{p}$ at $p>0.5$ \\
\middlehline
control & 0.019 & -- & -- & 0.017 \\
$[15,20]$, $\gamma=0.1$ & 0.254 & $-2.18$ & 2.96 & 0.020 \\
full, $\gamma=0.1$ (unmatched) & 0.911 & $-14.0$ & 131 & 1.33 (invalid) \\
full, $\gamma=0.01$ (matched $p$) & 0.202 & $-100$ & 203 & 0.00097 \\
\bottomhline
\end{tabular}
\end{table}

Table~\ref{tab:tier2} shows that in both full window cases the weights $\log o$  span hundreds of nats, and the resulting $\hat{p}$ disagrees by orders of magnitude with both the $[15,20]$-window estimate and the control, in the unmatched case even exceeding 1. The weighted sum in both full-trajectory cases is dominated by a single sample out of 704. Decomposing the variance of $\log o$ into the terms of eq.~\ref{eq:deltas}, we find that under full-trajectory guidance the variance is dominated ($\approx98\%$) by the reverse-path terms $\Delta_{latent}+\Delta_{denoiser}$, not by $\Delta_{guidance}$. We hypothesize that guiding at low $\sigma$, where the classifier is most sensitive to small perturbations, destabilizes the reverse-path divergence integral rather than the guidance term itself.

We conclude that the $[15,20]$ guidance window is required for an accurate importance-sampling estimate: guidance on the full trajectory gives a too-high-variance estimator for importance sampling in our case. 

\subsection{TC estimates under raw cBottle weights}\label{sec:app-raw-weights}
\begin{figure}
    \centering
    \includegraphics[width=0.9\linewidth]{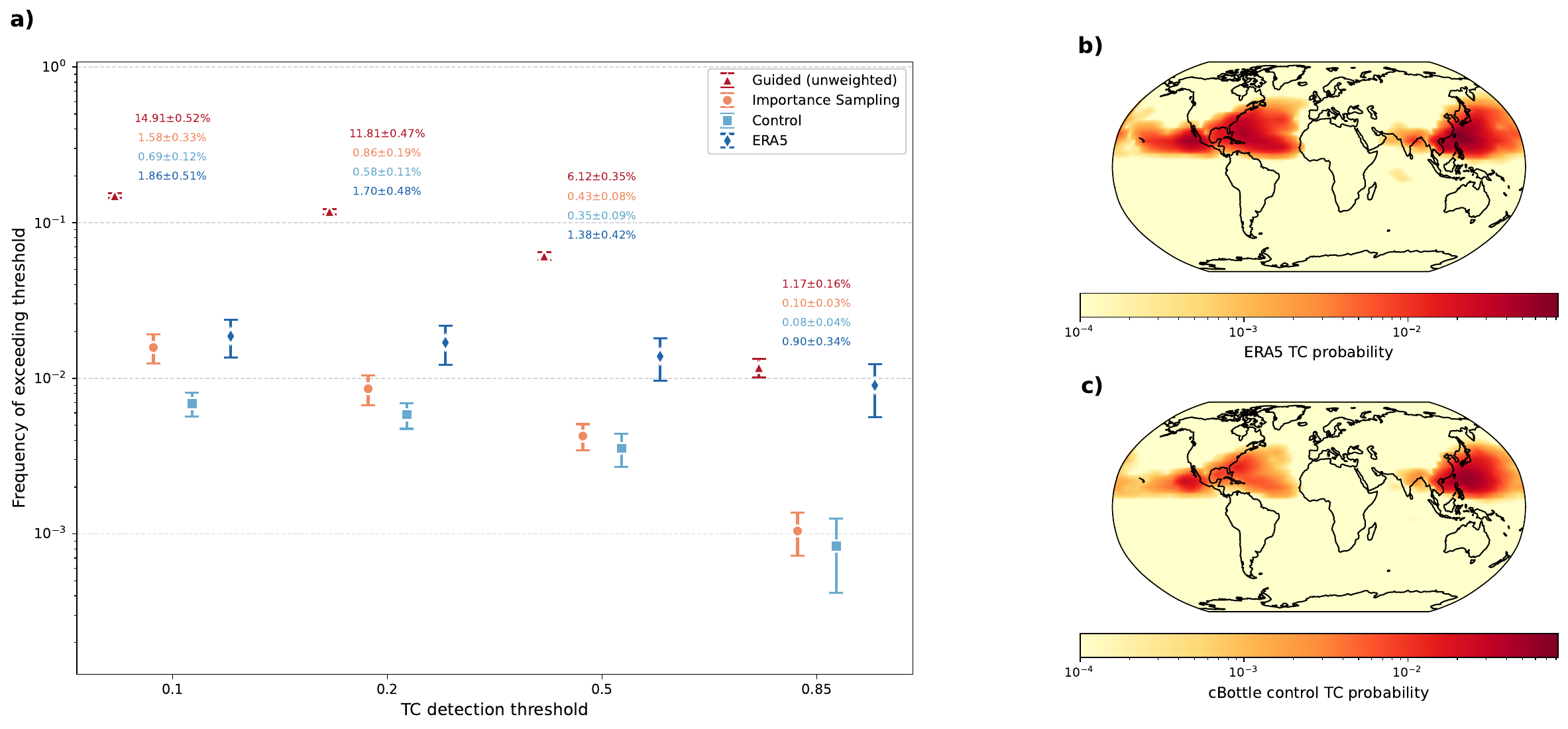}
    \caption{The same as Fig.~\ref{fig:is-tc} but using raw cBottle weights instead of EMA. Note that with the increased frequency of TCs in the control and IS estimates under EMA, higher TC detection thresholds are applied to arrive at similar frequencies.
    }
    \label{fig:is-tc-raw}
\end{figure}

In \cite{cbottle}, average TC frequencies are found to be too small. This is also shown here in Fig.~\ref{fig:is-tc-raw}, which is the same as Fig.~\ref{fig:is-tc} but for the published raw model weights (this is the figure from the first version of this manuscript). Control TC frequencies are lower than ERA5 at all thresholds. This difference becomes more pronounced in a) for higher TC detection thresholds, indicating that TCs are not only more frequent in ERA5 but also that they are more confidently detected than in the control. The plots in b) and c) show that this gap is larger in the North Atlantic and Eastern Pacific than in the Western Pacific.

\subsection{Physical TC intensity}\label{sec:tier3-intensity}
The classifier-probability thresholds used in Fig.~\ref{fig:is-tc} are a convenient but indirect proxy for TC intensity. We repeat the exceedance-frequency comparison using a physically interpretable quantity extracted from the same guided, control, and ERA5 states: the maximum 10\,m wind speed  in the HPX3 pixel containing Miami. At the HPX6 native inference resolution ($\sim$100\,km), these are resolution-limited proxies for storm intensity rather than best-track values.

\begin{figure}
    \centering
    \includegraphics[width=0.9\linewidth]{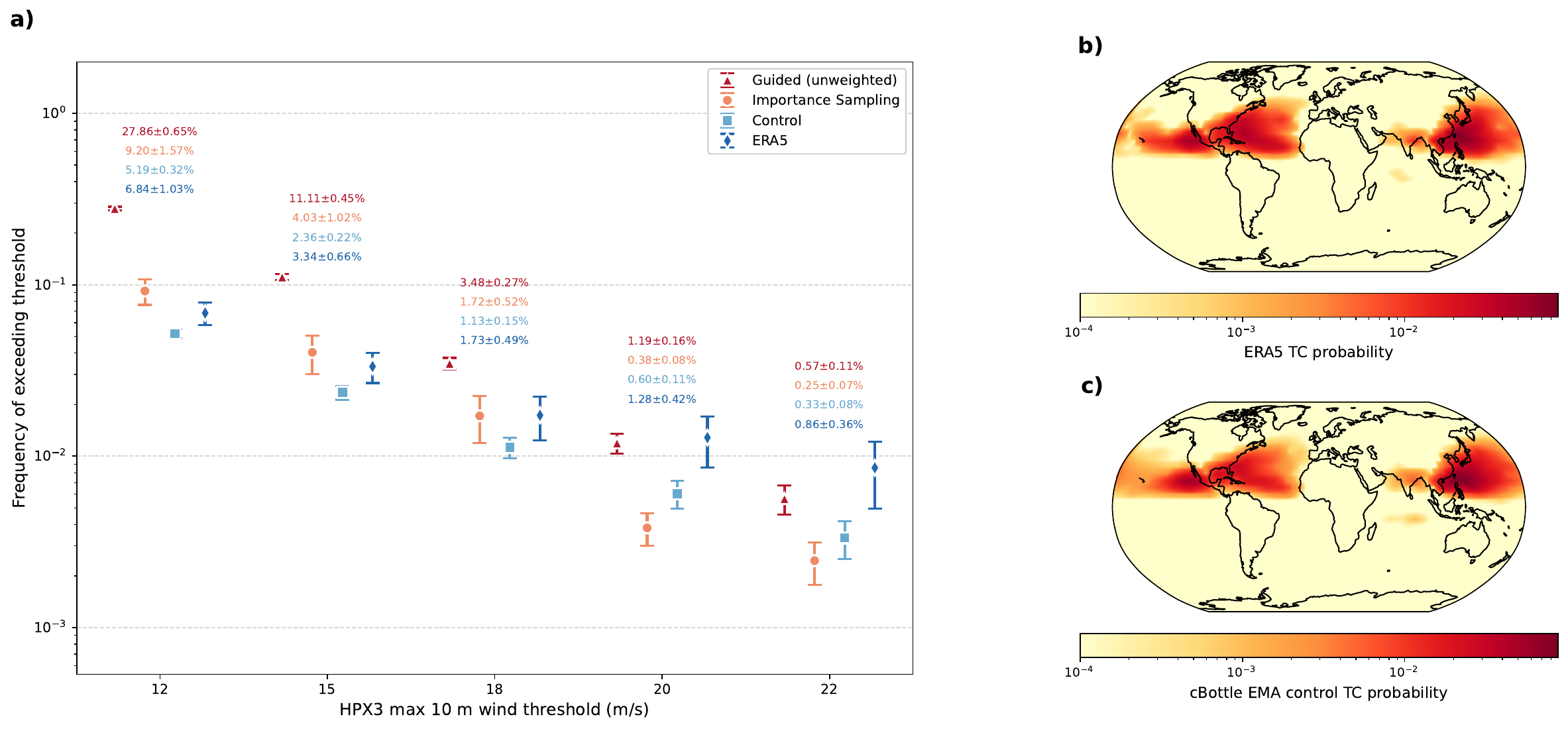}
    \caption{Analogue of Fig.~\ref{fig:is-tc}a), stratified by HPX3 maximum 10\,m wind-speed threshold (12, 15, 18, 20, 22\,m/s) instead of classifier probability, with ERA5 included.}
    \label{fig:tier3-wind}
\end{figure}

Figure~\ref{fig:tier3-wind} shows that at moderate wind-speed thresholds (12--18\,m/s) the importance-sampling estimate sits noticeably above the control, because the guided proposal targets classifier probability, not pixel-maximum wind speed directly: most of the importance-sampling weight at these thresholds comes from mixing in windy, non-TC states with order-1 odds ratios. At the higher thresholds (20--22\,m/s), where wind speed and classifier confidence are more tightly coupled, the importance-sampling estimate matches the control better and marginally  improves on its uncertainty, consistent with the classifier-threshold result of Fig.~\ref{fig:is-tc}. ERA5 frequencies are consistently higher than control's, which is an emulator bias that was not fixed by the use of EMA weights.

\begin{figure}
    \centering
    \includegraphics[width=0.9\linewidth]{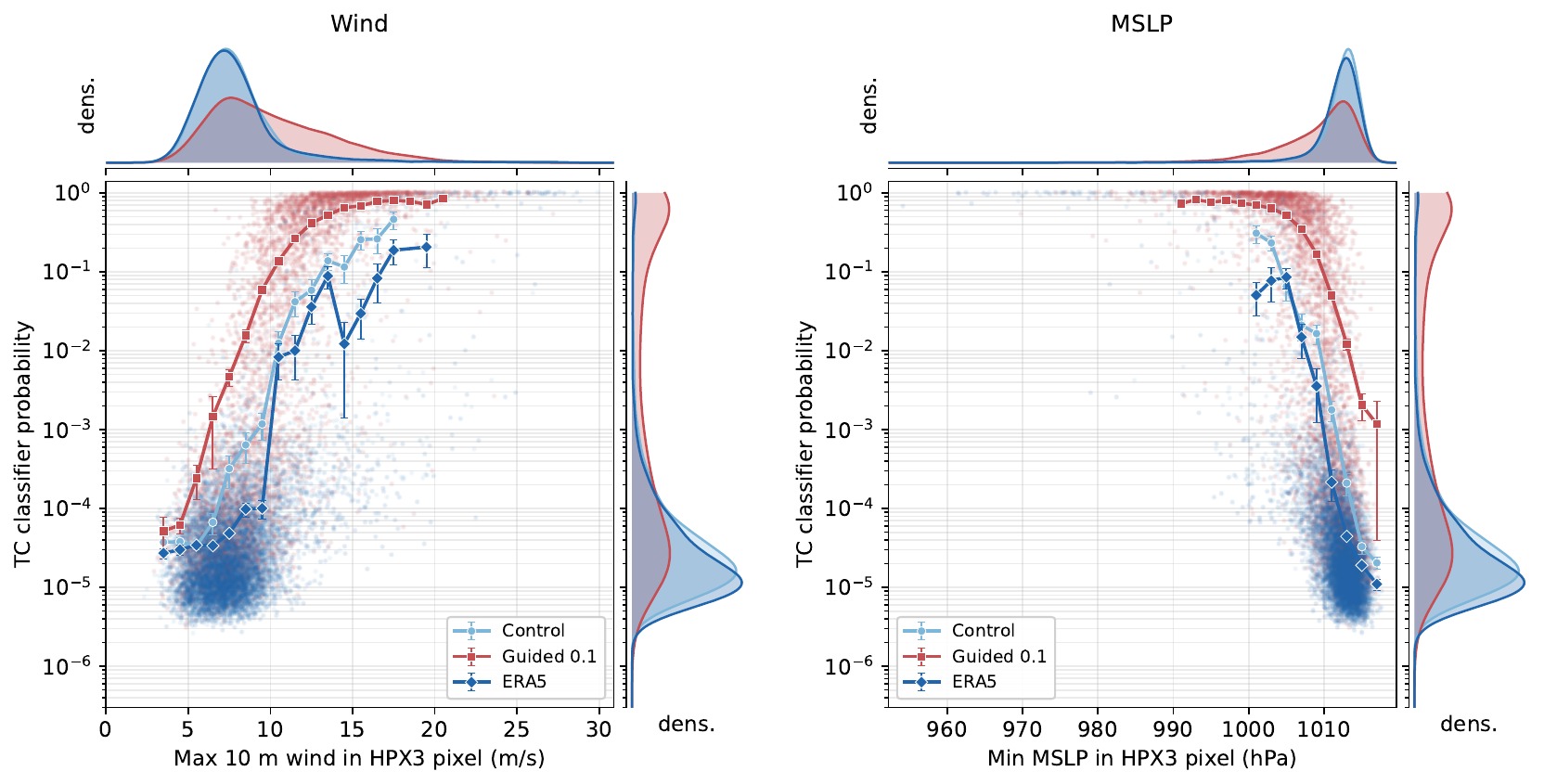}
    \caption{Classifier probability versus HPX3 maximum 10\,m wind speed and minimum mean sea-level pressure, with scatter, binned means, and marginal densities for guided, control, and ERA5 samples.}
    \label{fig:tier3-scatter}
\end{figure}

We note that TC detection probability is well correlated with the physical variables associated with a TC, as we show in Fig.~\ref{fig:tier3-scatter} for both maximum wind in the classified pixel and minimum pressure. However, guidance here has been performed on the TC probability to ensure a coherent feature across variables and meaningful TC spatial pattern. Performing the importance sampling estimate on one of the physical variables is therefore expected to yield higher-variance estimates than the results presented in Fig.~\ref{fig:is-tc}.

\subsection{Adaptive guidance scale}\label{sec:adaptive-guidance}
\cite{cbottle} use an adaptive guidance scale $\gamma = \hat{\gamma} \rho$, where $\hat{\gamma}$ is constant and
\begin{equation}
\rho = \frac{\lVert\frac{\mathbf{x}- \mathcal{D}_{\theta}(\mathbf{x}, \sigma)}{\sigma}\rVert^2}{\lVert\nabla_{\mathbf{x}} \mathcal{L}_{BCE}(\mathcal{C}(\mathbf{x}), y)\rVert^2}
\end{equation}
with the Frobenius norm $\lVert\cdot \rVert^2$. This ensures that at each step, the fraction of the norm of the update in the score direction and of that in the classifier direction is constant, and \cite{cbottle} found that without this rescaling, guidance did not produce TC samples. We find that setting the ratio $\rho=1$ (i.e. a constant guidance scale, section~\ref{sec:guidance-method}) is sufficient once guidance is additionally restricted to the $\sigma\in[15,20]$ window (section~\ref{sec:results}), since the norm-matching problem that motivated the adaptive scale is most severe at noise levels our reduced window excludes.

\subsection{Classifier skill and calibration versus noise level}\label{sec:tier4-classifier}
The TC classifier (section~\ref{sec:cbottle}) is trained jointly with the state denoiser as a \texttt{SongUNet} with an HPX3-resolution classification head, using binary labels derived from IBTrACS storm positions rasterized at HPX3, with a binary cross-entropy term (weight 0.1) added to the EDM denoising loss (eq.~\ref{eq:diffusionobj}) and trained over the same noise-level distribution $p_\sigma$ as the denoiser.

We evaluate classifier skill and calibration as a function of noise level $\sigma$ on held-out data: September 2018, 6-hourly ERA5 states (120 times) with IBTrACS HPX3 labels, disjoint from the training period ($<$2018), using the area under the ROC curve (AUC) restricted to tropical latitudes. 

\begin{figure}
    \centering
    \includegraphics[width=0.9\linewidth]{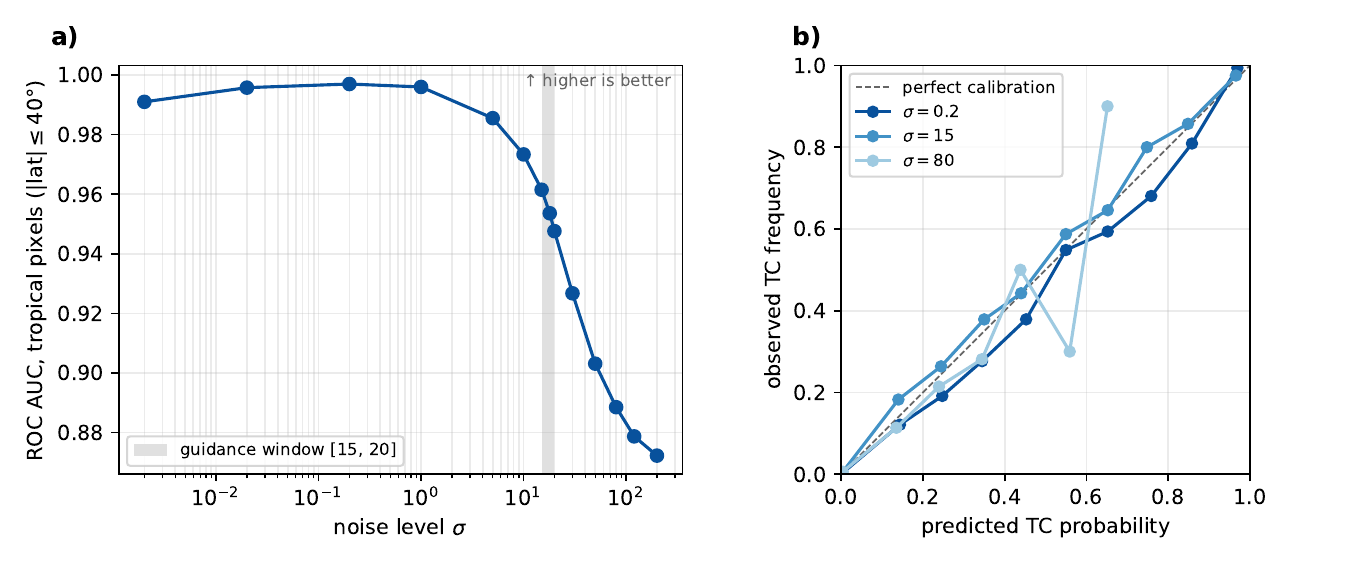}
    \caption{Classifier skill and calibration versus noise level, held-out September 2018 ERA5 with IBTrACS HPX3 labels. a) AUC (tropical latitudes) as a function of $\sigma$, with the guidance window $[15,20]$ shaded. b) Reliability diagram at $\sigma=0.2, 15, 80$, all pixels.}
    \label{fig:tier4-classifier}
\end{figure}

Classifier skill is highest at low $\sigma$ and decays with noise level (Fig.~\ref{fig:tier4-classifier}). The classifier nonetheless remains skillful and reasonably well calibrated within the guidance window (AUC $\approx0.95$--$0.96$ at $\sigma=15$--$20$).

As an independent check that classifier-flagged samples are physically consistent TCs, we additionally run TempestExtremes \texttt{DetectNodes} on remapped 1$^{\circ}$ fields. Base detection rates are 28.5\% for the guided archive, 1.75\% for the control, and 1.90\% for ERA5. Restricting to high-classifier-confidence samples, the fraction with an independent TempestExtremes hit at the Miami pixel is 92.6\% (guided), 89.1\% (control), and 98.5\% (ERA5) at classifier $p>0.5$, rising to 100\% for guided, control, and ERA5 at $p>0.98$ (38/38, 6/6, 5/5 samples respectively). This independent, classifier-free tracker therefore corroborates that high-classifier-probability guided samples are almost always detected as genuine cyclonic features, not classifier artifacts.

\subsection{Climate counterfactuals}     
\begin{figure}
    \centering
    \includegraphics[width=0.75\linewidth]{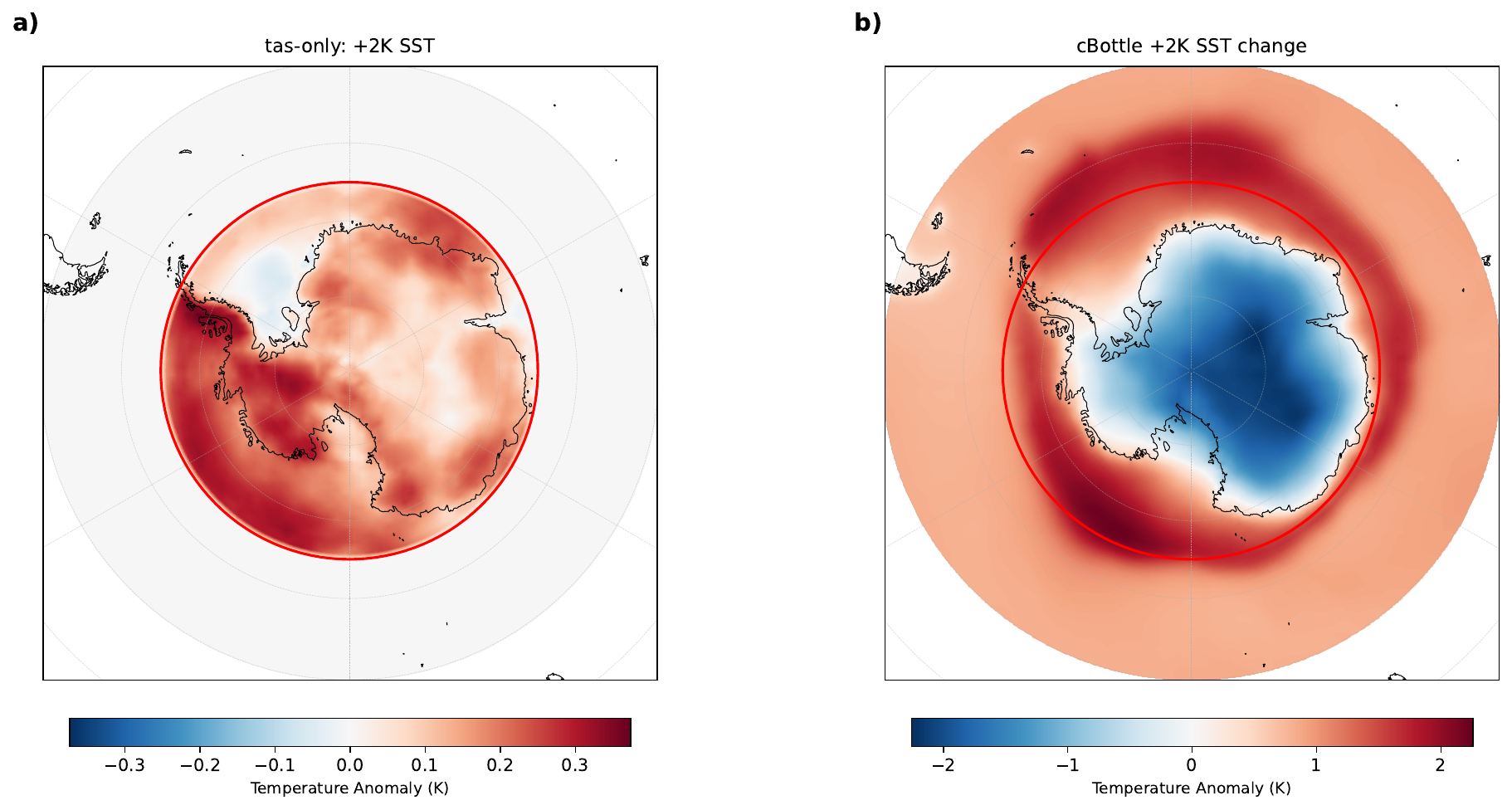}
    \caption{Comparing SST perturbation responses between the student model used here and the teacher model (cBottle).  }
    \label{fig:sst-pert}
\end{figure}

\noappendix       




\appendixfigures  

\appendixtables   


\authorcontribution{All authors contributed to the methodology, PM and NB developed the code for experiments, PM wrote the initial draft of the manuscript, all authors contributed to writing the manuscript.} 

\competinginterests{We declare no competing interests.} 


\begin{acknowledgements}
We would like to thank Akshay Subramaniam for helpful discussions and Elizabeth Barnes for feedback on the manuscript. 
\end{acknowledgements}

\bibliographystyle{copernicus}
\bibliography{references}

\end{document}